\documentclass[prl,aps,twocolumn,showpacs,preprintnumbers,amsmath,amssymb]{revtex4}

\usepackage{graphicx}
\usepackage{dcolumn}
\usepackage{bm}
\usepackage{color}

\begin{document}


\title{Quantum damping of Fermi-Pasta-Ulam revivals in ultracold Bose gases}
\author{Ippei Danshita$^{1}$}
\author{Rafael Hipolito$^{2}$}
\author{Vadim Oganesyan$^{2,3}$}
\author{Anatoli Polkovnikov$^{4}$}
\affiliation{
{$^1$Computational Condensed Matter Physics Laboratory, RIKEN, Wako, Saitama 351-0198, Japan}
\\
{$^2$The Graduate Center, CUNY, New York, NY 10016, USA}
\\
{$^3$Department of Engineering Science and Physics, College of Staten Island, CUNY, Staten Island, NY 10314, USA}
\\
{$^4$Department of Physics, Boston University, Boston, MA 02215, USA}
}

\date{\today}

\begin{abstract}
We propose an experimental scheme for studying the Fermi-Pasta-Ulam (FPU) phenomenon in a quantum mechanical regime using ultracold atoms. Specifically, we suggest and analyze a setup of one-dimensional Bose gases confined into an optical lattice. The strength of quantum fluctuations is controlled by tuning the number of atoms per lattice sites (filling factor). By simulating the real-time dynamics of the Bose-Hubbard model by means of the exact numerical method of time-evolving block decimation, we investigate the effects of quantum fluctuations on the FPU recurrence and show that strong quantum fluctuations cause significant damping of the FPU oscillation.

\end{abstract}

\maketitle
Systems of ultracold atomic gases in optical lattices have offered unique possibilities to study nonequilibrium dynamics of quantum many-body systems~\cite{bloch-08,polkovnikov-10}. Extreme cleanness and exquisite controllability of cold atom systems have led to experimental realization of various interesting nonequilibrium phenomena, such as the dynamics following the quench across the superfluid (SF)-Mott insulator (MI) transition~\cite{greiner-02}, the significant suppression of transport of one-dimensional (1D) lattice bosons~\cite{fertig-05, mun-07}, and the nonergodic dynamics of (nearly-)integrable 1D Bose gases~\cite{kinoshita-06}. In particular, the experiment of Kinoshita {\it et al.}~\cite{kinoshita-06} has shed light on the fundamental problem of the statistical mechanics, which is the relation between integrability and ergodicity, and stimulated renewed theoretical interest in this topic~(see Ref.~\cite{polkovnikov-10} and Refs. therein).

It is widely known through the celebrated numerical experiment by Fermi, Pasta, and Ulam (FPU) that non-integrable systems can also exhibit nonergodic dynamics under a certain condition~\cite{FPU-55}. In order to address the essential question whether the nonlinearity leads to ergodicity, FPU studied dynamics of 1D chains of classical nonlinear oscillators where initially only the first normal mode is excited. Contrary to their expectations, FPU found quasi-periodic recurrences of the initially excited mode, which are now called FPU recurrences. This discovery is marked as one of the most important milestones in the nonlinear physics because it largely stimulated the whole development of the chaos theory~\cite{campbell-05}. There have been multiple attempts to experimentally demonstrate FPU recurrences in several systems, including electrical networks~\cite{hirota-70}, double-plasma devices~\cite{ikezi-73}, deep water~\cite{lake-77}, magnetic films~\cite{scott-03, wu-07}, and optical fibers~\cite{simaeys-01}. However, such experiments usually suffer from the difficulty in making a physical system isolated from sources of energy dissipation that preclude the demonstration of complete recurrence~\cite{hirota-70, ikezi-73, lake-77,scott-03}. 

Theoretically, there does not appear to be a clear reason why FPU phenomenon, i.e. revivals of non-integrable many-body dynamics, should not be a general feature in low dimensional non-equilibrium physics, not particularly confined to the model studied by FPU. Indeed, in this Letter, we propose an experimental scheme for observing the FPU recurrences with ultracold Bose gases in optical lattices and present the numerical analysis of the corresponding dynamics. Ultracold atomic systems are especially suited for thorough studies of the FPU dynamics because of their unprecedented controllability and isolation. The central idea to achieve this goal is two-fold. First, when the filling factor is sufficiently large, the Bose-Hubbard (BH) model describing ultracold bosons in an optical lattice exhibits features of the FPU model, such as recurrences and energy localization in $q$-space. Secondly, in this regime, one can generate the initial state with the first normal mode excited prominently by means of phase imprinting. We show that dynamics of 1D Bose gases in optical lattices exhibit the FPU recurrences in the classical (Gross-Pitaevskii) limit. Moreover, such setup has another advantage allowing one to control the strength of quantum fluctuations through varying the filling factor, the interatomic interaction, and the lattice depth. To our knowledge this is the first experimentally feasible proposal to study the FPU phenomenon in the quantum regime. We also analyze dynamics in the suggested setup using the time-evolving block decimation (TEBD) method~\cite{vidal-04}. We find that the quantum fluctuations cause damping of the FPU recurrences. Such damping can be reduced by increasing the filling factor such that in the classical limit we find nearly complete revivals characteristic for the original FPU problem.

We consider a system of $N$ bosons in a 1D optical lattice. Assuming that the lattice is sufficiently deep, the system can be described by the 1D BH Hamiltonian~\cite{fisher-89}:
\begin{eqnarray}
\hat{H} = -J\sum_{j=1}^{L}( \hat{b}^{\dagger}_j \hat{b}_{j+1} + {\rm h.c.})
            + \frac{U}{2}\sum_{j=1}^{L} \hat{b}^{\dagger}_j \hat{b}^{\dagger}_j \hat{b}_j \hat{b}_j.
\label{eq:BHH}
\end{eqnarray}
where the field operator $\hat{b}^{\dagger}_j$ ($\hat{b}_j$) creates (annihilates) a boson on the $j$-th site, $J$ is the hopping energy, and $U$ is the onsite interaction. To make closer connection with the FPU's original model, we assume the absence of a harmonic trap that is used in standard experiments. This assumption can be valid when one uses a box-shaped trap~\cite{meyrath-05} or a ring optical lattice~\cite{henderson-09}. We henceforth focus
on the case of a ring geometry for theoretical simplicity taking the periodic boundary conditions in Eq.~(\ref{eq:BHH}).

The strength of quantum fluctuations is characterized by the effective Planck's constant $h_{\rm e} \equiv \sqrt{U/(\nu J)}$, where $\nu=N/L$ is the filling factor. Physically $h_{\rm e}$ controls proximity of the system to the Mott insulator regime with the transition occurring at $h_{\rm e}\simeq 1.5$ for integer large $\nu$. $h_{\rm e}$ can be experimentally varied in a broad range~\cite{polkovnikov-05,danshita-10}. We note that the nonlinearity in the system is controlled by another parameter $\kappa= U\nu/J$. When this parameter is large, the number fluctuations in the system are small and the Hamiltonian (\ref{eq:BHH}) can be mapped into the quantum rotor model~\cite{polkovnikov-05}.

Let us first consider the classical, Gross-Pitaevskii, limit ($h_{\rm e}\rightarrow 0$) and identify parameter regions where FPU recurrences occur. In this limit, one can safely neglect quantum fluctuations and replace the field operator $b_j(t)$ with a classical field $\Psi_j(t)$. Carrying out the replacement in Eq.~(\ref{eq:BHH}), we obtain the classical Hamiltonian:
\begin{eqnarray}
\mathcal{H} = -J\sum_{j=1}^{L}( \Psi^{\ast}_j \Psi_{j+1} + {\rm c.c.})
            + \frac{U}{2}\sum_{j=1}^{L} |\Psi_{j}|^4.
\end{eqnarray}
The dynamics of the system can be described by the discrete nonlinear Schr\"odinger equation (DNLSE), or equivalently discrete Gross-Pitaevskii equation~\cite{trombettoni-01}:
\begin{eqnarray}
i\hbar \frac{d \Psi_j}{d t} = \frac{\delta \mathcal{H}}{\delta \Psi^{\ast}_j}
= - J (\Psi_{j-1} + \Psi_{j+1}) + U |\Psi_j|^2 \Psi_j.
\label{eq:dnlse}
\end{eqnarray}
It is clear that in this classical limit both static and dynamic properties of the DNLSE (after trivial time rescaling by $J$) are determined by the parameter $\kappa = U\nu / J$ introduced earlier.  In the rotor regime $\kappa \gg 1$, DNLSE can be mapped into
\begin{eqnarray}
\frac{d^2 \phi_j}{dt^2} \!=\!
\frac{2\nu J U}{\hbar^2}\left( \sin\varphi_{j-1,j} + \sin\varphi_{j+1,j} \right), \,
\frac{d\phi_j}{dt} \!=\! - \frac{U}{\hbar} n_j,
\label{eq:crotor}
\end{eqnarray}
where $\phi_j$ and $n_j$ are the phase and the density of the condensate defined by $\Psi_j = \sqrt{n_j}e^{i\phi_j}$ and $\varphi_{j,l} \equiv \phi_j - \phi_l$. Notice that the condition $\kappa \gg 1$ is compatible with $h_{\rm e}\ll 1$ at large filling factors $\nu \gg 1$.
DNLSE in the rotor regime reduces to the quartic FPU model when the nonlinearity is small: $|\phi_j-\phi_{j+1}|\ll 1$, so that one can expand the sinusoidal terms as $\sin\varphi \simeq \varphi - \varphi^3/6$. However, there are also some differences; for example, the DNLSE conserves the number of particles, $\sum_j |\Psi_j|^2 = N$, while there is no such conservation in the latter models. In addition at large but finite $\kappa$ there is a small quadratic nonlinearity present in Eq.~(\ref{eq:crotor}) which is equivalent to an additional small cubic nonlinearity in the FPU model. Indeed,  we work in the regime where cubic nonlinearity is not negligible.  We stress that this is perfectly fine for studying FPU phenomena since the only necessary condition is for $\kappa$ to be sufficiently large, and we do not need to be strictly in the rotor limit.

\begin{figure}[tb]
\includegraphics[scale=0.45]{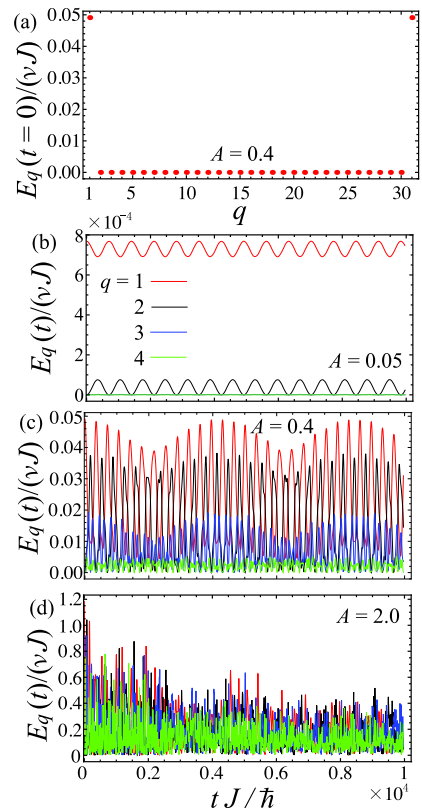}
\caption{\label{fig:GP}
(color online) (a) The energy of all the normal modes at $t=0$, where $A=0.4$.
(b)-(d)Time evolution of the energy of the first four modes $A=0.05$ (b), $0.4$ (c), and $2.0$ (d).
We set $L=32$ and $\kappa = 20$.
}
\end{figure}

In order to induce FPU recurrence, we consider the following initial condition,
\begin{eqnarray}
\phi_j(t=0) = A \sin\left(2\pi j/L\right), \, n_j(t=0) = \nu,
\label{eq:PI}
\end{eqnarray}
which can be generated with the use of phase-imprinting techniques~\cite{burger-1999}. The amplitude $A$ is related to the energy $E_{\Delta}=\mathcal{H}(t=0)-\mathcal{H}_{\rm gs}$ injected by this phase imprint as $E_{\Delta} \simeq 2 N J (\pi A/L)^2$, where $\mathcal{H}_{\rm gs}$ is the classical ground-state energy.  Fig.~\ref{fig:GP}(a) shows the initial energy distribution of the normal modes $E_q(t=0)$ for the chosen initial condition~(\ref{eq:PI}) for $L=32$ and $\kappa=20$. The normal mode energies are given by
\begin{eqnarray}
E_q = 4\nu J \sin^2\left(\frac{\pi}{L}q\right)|\tilde{\phi}_q|^2  + \frac{U}{2}|\tilde{n}_q|^2.
\label{eq:modeE}
\end{eqnarray}
In Eq.~(\ref{eq:modeE}),  $\tilde{\phi}_q$ and $\tilde{n}_q$ are the Fourier transforms of the phase $\phi_j$ and the density $n_j$ given by
%
$
\tilde{\phi}_q = L^{-1/2} \sum_j \phi_j e^{- i 2\pi q j/L},
\,
\tilde{n}_q = L^{-1/2} \sum_j n_j e^{- i 2\pi q j/L},
$
%
and the mode number $q$ is integer satisfying $1\leq q \leq L-1$.  
In Fig.~\ref{fig:GP}(a), it is clear that only the normal modes with the lowest frequency ($q=1$ and $L-1$) are prominently excited in the initial state as in the original work of FPU.  Therefore, the FPU recurrences are expected to occur in the dynamics subjected to the initial phase imprint.

In Figs.~\ref{fig:GP}(b)-(d), we show time evolution of the energy of the first four modes for several values of the amplitude. These results are obtained by numerically solving DNLSE~(\ref{eq:dnlse}). When the amplitude $A$ is small as in Fig.~\ref{fig:GP}(b), the energy of the initially excited mode is almost constant because the coupling among the normal modes is too weak to induce the mode mixing needed for the FPU dynamics. As $A$ increases, the energy spreads to other modes but still remains confined to a few of them. In this regime one observes repetitive recurrences of the first mode (Fig.~\ref{fig:GP}(c)). Both long-time revivals and even longer time super-revivals, where nearly $100\%$ of the energy returns to the first mode, characteristic for the FPU dynamics~\cite{campbell-05} are clearly present. Finally as $A$ increases even further and exceeds a certain threshold, the dynamics becomes chaotic rapidly reaching equipartition of the energy between all the modes (Fig.~\ref{fig:GP}(d)). Such a threshold for ergodic behavior has been previously found in many other models including the quartic FPU model~\cite{izrailev-66,livi-85}, the nonlinear Schr\"odinger equation with open boundaries~\cite{villain-00}, and DNLSE at smaller filling~\cite{cassidy-09}.

\begin{figure}[tb]
\includegraphics[scale=0.48]{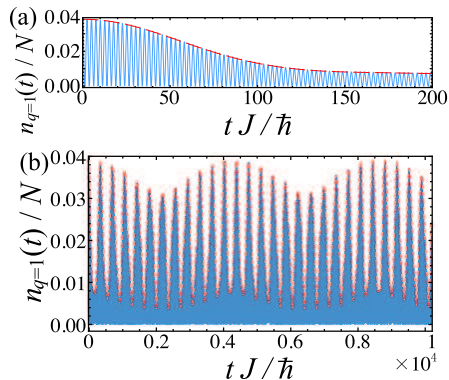}
\caption{\label{fig:mom}
(color online) Time evolution of the momentum occupation $n_q(t)$ for $q=1$ (blue solid line). The parameters of the system are the same as in Fig.~\ref{fig:GP}(c). The red dashed line represents $C\times E_{q=1}(t)$, where the constant $C$ is chosen to agree with the envelope of the oscillation of $n_{q=1}(t)$. The graphs (a) and (b) show short and long time dynamics respectively.
}
\end{figure}

The temporal behavior of the mode energies shows close analogy between BH and FPU models
. However these mode energies are not easily observable in cold atom experiments. It is thus much more natural to analyze time evolution of the quasi-momentum occupation: $n_q = L^{-1} \sum_{j,l}\Psi_j^{\ast} \Psi_l e^{i2\pi q(j-l)/L}$ , which is a standard observable~\cite{bloch-08}. In fig.~\ref{fig:mom}(a) we plot $n_{q=1}(t)$. Clearly the momentum distribution oscillates at much larger frequency $\omega\approx\sqrt{2\nu J U}\pi/(L\hbar)$ compared to frequency of the energy oscillations. These oscillations are expected even in the absence of the coupling between normal modes. From fig.~\ref{fig:mom}(b) it is obvious that the envelope of these fast oscillations exhibits the FPU recurrences and thus the quasi-momentum distribution can serve as a good observable for detecting the FPU dynamics.

\begin{figure}[tb]
\includegraphics[scale=0.42]{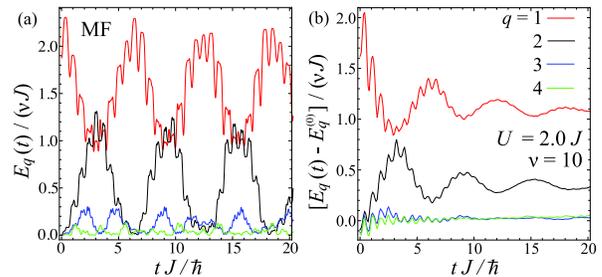}
\caption{\label{fig:CQ}
(color online) Time evolution of the energy $E_{q}$ of the first four modes, where $L=8$ and $U\nu/J = 20$.
(a) The classical dynamics based on DNLSE.
(b) The quantum dynamics calculated by applying TEBD to the Bose-Hubbard model.
We set $\nu=10$, which means that the effective Planck's constant $h_{\rm e}= \sqrt{0.2}\simeq 0.447$.
}
\end{figure}
Having confirmed that the FPU recurrences occur below the equilibration threshold in the classical limit, we now consider quantum dynamics of our system. For this purpose we apply the quasi-exact numerical method of TEBD~\cite{vidal-04} with periodic boundary conditions~\cite{danshita-09} to the BH model. For efficient TEBD simulations for large filling factor ($\nu \gg 1$) we reduce the dimension of the local Hibert space by introducing a lower bound of the particle occupation number in addition to an upper bound~\cite{danshita-10}. An initial state for the quantum FPU dynamics is generated by preparing the system in the ground state of the Hamiltonian~(\ref{eq:BHH}) via the imaginary time propagation and then applying the phase imprint operator, $\hat{P}=\prod_{j=1}^L e^{-i\theta_j \hat{b}_j^{\dagger}\hat{b}_j}$, to the ground state at $t=0$, where $\theta_j =A\sin(2\pi j/L)$. As in the classical case, we calculate the time evolution of the mode energies $E_q(t)$ additionally subtracting the contribution from the depletion $E_q^{(0)}$ which is equal to the energy of the corresponding mode in the ground state.

\begin{figure}[tb]
\includegraphics[scale=0.58]{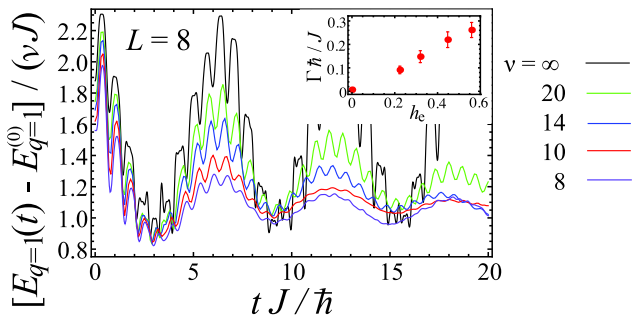}
\caption{\label{fig:nus}
(color online) Time evolution of the energy $E_{q=1}$ of the first mode for $\nu=\infty$ (DNLSE), $20, 14, 10$, and $8$ where $L=8$ and $U\nu/J = 20$.
The inset depicts the damping rate $\Gamma$ of the FPU oscillation as a function of the effective Planck's constant $h_{\rm e} \equiv \sqrt{U/(\nu J)}$.
}
\end{figure}

In Fig.~\ref{fig:CQ}, we show time evolution of the mode energies both for the classical (a) and the quantum (b) dynamics. In the quantum case we use $\nu=10$ corresponding to $h_{\rm e} \simeq 0.447$. In both cases $\kappa = 20$ and $L=8$. Note that $L=8$ is a minimum possible size required to observe the FPU dynamics. For smaller systems the frequency of the FPU oscillation is too close to that of the lowest-energy excitation. For even this relatively small system the Hilbert space is too large to allow for exact diagonalization. From Fig~\ref{fig:CQ}(b) it is clear that the FPU oscillations are significantly damped in the quantum case. To corroborate that this damping is due to quantum fluctuations,  we vary the filling factor $\nu$ while fixing $\kappa = 20$, and analyze how the damping depends on the effective Planck's constant $h_{\rm e}$. In Fig.~\ref{fig:nus} we show the time evolution of the first mode energies for several values of $\nu$. Clearly the damping becomes stronger at smaller $\nu$, i.e. larger $h_{\rm e}$. We extract the damping rate $\Gamma$ by fitting $E_{q=1}(t)$ to a function $f(t) = a e^{-\Gamma t}\cos(bt) + c$, where $a$, $b$, $c$, and $\Gamma$ are free parameters. In the inset of Fig.~\ref{fig:nus}, we plot the extracted damping rate versus $h_{\rm e}$. When the filling $\nu$ becomes even smaller so that the system approaches the Mott insulator phase we find that the mode energies start exhibiting new quantum revivals which are not related to the FPU physics, but rather to the gradual emergence of the particle-hole excitations.

It is worth noting that the damping of the FPU oscillation does not result in equilibration within the timescale of our TEBD calculations. This implies that quantum fluctuations do not necessarily make the FPU dynamics ergodic. Instead, the system seems to approach a certain metastable state, and it is interesting to seek a statistical mechanics describing such a metastable state. We will address this issue in detail elsewhere~\cite{hipolito-10}.

One possible qualitative explanation for the damping can be made from the perspective of the decay of mater-wave solitons due to quantum fluctuations. On the basis of the soliton concept, the classical FPU recurrences are interpreted as follows~\cite{zabusky-65}. The initially prepared wave with the sine form splits into a number of solitons with different velocities. Since each soliton is stable during time evolution, the solitons come back very close to the initial configuration after colliding many times, producing the recurrences.  In the quantum case, on the contrary, it was argued that the solitons are unstable~\cite{mishmash-09,martin-10} and that the soliton lifetime decreases with strengthening quantum fluctuations~\cite{mishmash-09}. This soliton decay may be responsible for the damping of the FPU oscillations.

In conclusion, we analyzed the dynamics of one dimensional bosons in optical lattices and demonstrated that under initial phase imprint at high filling factors FPU recurrences occur in this system. We numerically studied time evolution of this system to reveal quantum effects on FPU recurrences. We showed that the damping of the FPU oscillations becomes more significant with increasing the strength of quantum fluctuations. We emphasize that the controllability of quantum fluctuations in ultracold atom systems can open up a new direction of the study of the FPU phenomenon.

\begin{acknowledgments}
I. D. was supported by KAKENHI (22840051) from JSPS. I. D. thanks Boston University visitors program for hospitality and T. Kinoshita for stimulating discussions. R.H. was supported by  NSF CUNY AGEP Award 0450360. V.O. was supported by NSF (DMR-0955714). A.P. was supported by NSF (DMR-0907039), AFOSR FA9550-10-1-0110, and
Sloan Foundation. The computation in this work was partially done using the RIKEN Cluster of Clusters facility.
\end{acknowledgments}



\begin{thebibliography}{99}
\bibitem{bloch-08}
I. Bloch, J. Dalibard, and W. Zwerger,
Rev. Mod. Phys. {\bf 80}, 885 (2008).

\bibitem{polkovnikov-10}
A. Polkovnikov, K. Sengupta, A. Silva, and M. Vengalattore,
arXiv:1007.5331v1.

\bibitem{greiner-02}
M. Greiner {\it et al.},
Nature (London) {\bf 419}, 51 (2002).

\bibitem{fertig-05}
C. D. Fertig {\it et al.},
Phys. Rev. Lett. {\bf 94}, 120403 (2005).

\bibitem{mun-07}
J. Mun {\it et al.},
Phys. Rev. Lett. {\bf 99}, 150604 (2007).

\bibitem{kinoshita-06}
T. Kinoshita {\it et al.},
Nature (London) {\bf 440}, 900 (2006).

\bibitem{FPU-55}
E. Fermi, J. Pasta, and S. Ulam,
Los Alamos Science Laboratory Report No. LA-1940 (1955).

\bibitem{campbell-05}
D. K. Campbell, P. Rosenau, and G. M. Zaslavsky,
Chaos {\bf 15}, 015101 (2005);
G. P. Berman and F. M. Izrailev,
{\it ibid.} {\bf 15}, 015104 (2005).


\bibitem{hirota-70}
R. Hirota and K. Suzuki,
J. Phys. Soc. Jpn. {\bf 28}, 1366 (1970).

\bibitem{ikezi-73}
H. Ikezi,
Phys. Fluids 16, 1668 (1973).

\bibitem{lake-77}
B. M. Lake {\it et al.},
J. Fluid Mech. {\bf 83}, 49 (1977).

\bibitem{scott-03}
M. M. Scott {\it et al.},
J. Appl. Phys. {\bf 94}, 5877 (2003).

\bibitem{wu-07}
M. Wu and C. E. Patton,
Phys. Rev. Lett. {\bf 98}, 047202 (2007).

\bibitem{simaeys-01}
G. Van Simaeys {\it et al.},
Phys. Rev. Lett. {\bf 87}, 033902 (2001).

\bibitem{vidal-04}
G. Vidal,
Phys. Rev. Lett. {\bf 93}, 040502 (2004).

\bibitem{fisher-89}
M. P. A. Fisher {\it et al.},
Phys. Rev. B {\bf 40}, 546 (1989).

\bibitem{meyrath-05}
T. P. Meyrath {\it et al.},
Phys. Rev. A {\bf 71}, 041604(R) (2005).

\bibitem{henderson-09}
K. Henderson {\it et al.},
New. J. Phys. 11, 043030 (2009).

\bibitem{polkovnikov-05}
A. Polkovnikov {\it et al.},
Phys. Rev. A {\bf 71}, 063613 (2005).

\bibitem{danshita-10}
I. Danshita and A. Polkovnikov,
Phys. Rev. B {\bf 82}, 094304 (2010).

\bibitem{trombettoni-01}
A. Trombettoni and A. Smerzi,
Phys. Rev. Lett. {\bf 86}, 2353 (2001).

\bibitem{burger-1999}
S. Burger {\it et al.},
Phys. Rev. Lett. {\bf 83}, 5198 (1999).

\bibitem{izrailev-66}
F. M. Izrailev and B. V. Chirikov,
Sov. Phys. Dokl. {\bf 11}, 30 (1966).

\bibitem{livi-85}
R. Livi {\it et al.},
Phys. Rev. A 31, 1039 (1985).

\bibitem{villain-00}
P. Villain and M. Lewenstein,
Phys. Rev. A {\bf 62}, 043601 (2000).

\bibitem{cassidy-09}
A. C. Cassidy {\it et al.},
Phys. Rev. Lett. {\bf 102}, 025302 (2009).

\bibitem{danshita-09}
I. Danshita and P. Naidon,
Phys. Rev. A {\bf 79}, 043601 (2009).

\bibitem{hipolito-10}
R. Hipolito, I. Danshita, V. Oganesyan, and A. Polkovnikov,
in preparation.

\bibitem{zabusky-65}
N. J. Zabusky and M. D. Kruskal,
Phys. Rev. Lett. {\bf 15}, 240 (1965).

\bibitem{mishmash-09}
R. V. Mishmash and L. D. Carr,
Phys. Rev. Lett. {\bf 103}, 140403 (2009);
R. V. Mishmash {\it et al.},
Phys. Rev. A {\bf 80}, 053612 (2009).

\bibitem{martin-10}
A. D. Martin and J. Ruostekoski,
Phys. Rev. Lett. {\bf 104}, 194102 (2010).

\end{thebibliography}
\end{document}